\documentclass[11pt,twoside]{article}
\usepackage{graphicx,rotating,cite,epsfig,amstex}

\newcommand{\jpsi}      {\mbox{J\kern-0.05em /\kern-0.05em$\psi$}}
\newcommand{\psip}      {\mbox{$\psi^\prime$}}
\newcommand{\rs}        {\mbox{$\sqrt{s}$}}
\newcommand{\dy}        {Drell-Yan}
\newcommand{\ddbar}     {$D\overline{D}$}
\newcommand{\dbar}      {$\overline{D}$}
\newcommand{\dd}        {\mathrm{d}}
\newcommand{\pt}        {$p_{\mathrm{T}}$}

\newcommand{\xf}        {$x_{\mathrm{F}}$}
\newcommand{\Pythia}    {\textsc{Pythia}}
\newcommand{\mkt}       {\mbox{$\langle k_{\rm T}^2 \rangle$}}

\def    \ee     {\mbox{$e^+e^-$}}
\def    \mm     {\mbox{$\mu^+\mu^-$}}

\def    \pim    {\mbox{$\pi^-$}}

\def    \?      {\mbox{$\spadesuit$}}

\def    \gev    {\mbox{~GeV}}
\def    \gevc   {\mbox{~GeV/$c$}}
\def    \gevcc  {\mbox{~GeV/$c^2$}}
\def    \pp     {\mbox{pp}}
\def    \pa     {\mbox{pA}}

\def    \cc     {\mbox{$c \overline {c}$}}
\def    \PbPb   {\mbox{~Pb+Pb}}
\newcommand \beq{\begin{eqnarray}}
\newcommand \eeq{\end{eqnarray}}

\begin{document}

\begingroup
\thispagestyle{empty}
\baselineskip=14pt
\parskip 0pt plus 5pt

\begin{center}
{\large EUROPEAN ORGANIZATION FOR NUCLEAR RESEARCH}
\end{center}

\begin{flushright}
CERN--PPE\,/\,97--65--Rev\\
August 5, 1997
\end{flushright}
\bigskip

\begin{center}
\Large
\textbf{Open charm contribution to  dilepton spectra}
\textbf{produced in nuclear collisions at SPS energies}
\end{center}

\bigskip\bigskip
\begin{center}
P.~Braun-Munzinger, D.~Mi\'skowiec\\
\emph{GSI, Darmstadt, Germany}\\
\bigskip
A.~Drees\\
\emph{Universit\"at Heidelberg, Germany}\\
\bigskip
C.~Louren\c{c}o\\
\emph{CERN, Geneva, Switzerland}\\
\end{center}
\bigskip\bigskip\bigskip

\begingroup
\leftskip=0.4cm
\rightskip=0.4cm
\parindent=0.pt
\begin{center}
ABSTRACT
\end{center}
Measurements of open charm hadro-production from CERN and Fermilab
experiments are reviewed, with particular emphasis on the absolute
cross sections and on their A and $\sqrt{s}$ dependences.
Differential \pt\ and \xf\ cross sections calculated with the \Pythia\
event generator are found to be in reasonable agreement with recent
data.  The calculations are scaled to nucleus-nucleus collisions and
the expected lepton pair yield is deduced. The charm contribution
to the low mass dilepton continuum observed by the CERES experiment is
found to be negligible. In particular, it is shown that the observed low
mass dilepton excess in S-Au collisions cannot be explained by charm
enhancement.
\endgroup

\vfill
\begin{center}
Accepted for publication in Z.~Phys.~C
\end{center}

\endgroup


\newpage
\pagenumbering{arabic}
\setcounter{page}{2}

\section{Introduction}

In order to understand whether the data collected in nucleus-nucleus
collisions signal new physics such as quark-gluon plasma formation,
there are essentially two possible procedures.
The first one consists in using Monte Carlo event generators
to calculate what should happen in  conventional scenarios.
A significant difference between  prediction and  experiment would
indicate new physics.  The standard event generator codes, however,
simulate nucleus-nucleus collisions only in what concerns soft
processes.  Physics results based on conventional  hard processes in
nucleus-nucleus collisions, on the other hand, can be predicted by
extrapolation from \pp\ and \pa\ data.  This second method has been
used in the study of high mass dimuon production.  The present
understanding of \jpsi, \psip\ and \dy\ production in \pa\
interactions, from data collected at CERN and at Fermilab, allows to
estimate the behavior expected in S and Pb induced
collisions~\cite{clqm96}.

Open charm production is a hard process and the presently available
nucleus-nucleus event generators cannot be used to estimate production
yields or contributions to the dilepton spectra measured at SPS
energies.  The SPS heavy ion collaborations NA38 and HELIOS-3 have
used the \Pythia\ Monte Carlo event generator \cite{pythia} to calculate charm
production on the \pp\ level and used a  scaling which is linear with
the nuclear mass number A to compare the
estimated values to the data collected with  sulphur
beams~\cite{clqm93,Mazzoni}.
An initial attempt to assess the charm decay contribution to
dilepton spectra at higher masses has been presented by A.~Shor~\cite{sho91}.
In this paper we use \Pythia\ to evaluate whether the low mass
electron pair excess observed by CERES \cite{aga95}
might be explained by charm enhancement
in heavy ion collisions.

The theoretical understanding of the features seen in the charm data
for \pp\ and \pa\ collisions has
made substantial progress in  recent years but is still facing
some difficulties (see~\cite{Mangano} for a recent review).
The availability of NLO calculations has not eliminated the need
for uncomfortably large values for the intrinsic transverse momentum
$k_{\mathrm{T}}$ 
of the incoming partons (the `$k_{\mathrm{T}}$~kick').
Adding together the uncertainty in c-quark mass ($m_c$ ranging from 1.2 to
1.8\gevcc), values of \mkt\ that can go up to 2~GeV$^2$/$c^2$,
doubts on which fragmentation scheme better describes the hadronization
process, and the ignorance of higher order 
effects, especially at low transverse momentum,
we get the feeling that the theoretical
understanding is still somewhat unclear. For a detailed discussion of
these theoretical uncertainties including a complete list of references
see~\cite{Mangano}. 
It is not our intention to clarify this situation.  Rather we wish 
to evaluate whether a \Pythia\ based model is able to
describe reasonably well the available data on open charm production.
If so, we can use it as a good basis to predict,
in the absence of new physics,
charm production in nucleus-nucleus collisions and to
estimate the charm contribution to dilepton spectra.

In Section 2 we compare the results provided by \Pythia\ with the
recent FNAL and CERN data on open charm hadro-production, with
particular emphasis on the energy dependence of the charm cross
section, on the distributions in transverse momentum \pt\ and the
Feynman x variable \xf\,  and on the correlations
between charmed hadrons.  We then compute the lepton pair yield from
simultaneous semileptonic decays of $D$ and \dbar\ mesons and
$\Lambda_c$ baryons and compare
it to the experimentally detected inclusive dilepton yield in p-p and
p-A collisions. Scaling the results to nucleus-nucleus collisions and
comparing them to data as well as the possibility of enhanced charm
production in these collisions are discussed in Section 4.

\section{Charm production in \pp\ and \pa\ collisions}

\begin{figure}[t]
\centering
\resizebox{\textwidth}{!}{\includegraphics{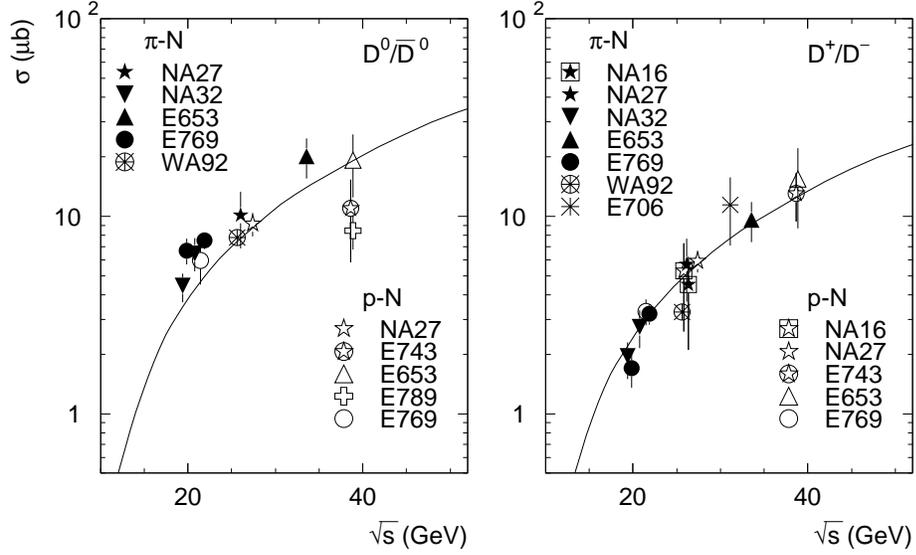}}
\caption{{} Cross section for forward (\xf$>$0)
  $D^0,\overline{D}^0$ (left) and $D^+,D^-$
  (right) production in proton and pion induced reactions,
  as a function of the center of mass  energy.
  The data were taken from NA16 \cite{na16}, NA27 \cite{na27},
  E743 \cite{e743}, E653 \cite{e653}, E789 \cite{e789},
  E769 \cite{e769}, NA32 \cite{na32}, WA92 \cite{wa92}, and E706 \cite{e706}.
  The solid lines represent the \pp\ \Pythia\ calculations,
  up-scaled by a factor of 3.5 (left panel) and 7 (right panel).
}
\label{fig:xsection}
\end{figure}

The recent Fermilab and CERN experiments on the production of $D$ mesons
in proton and pion induced collisions
provide a rather detailed picture of charm hadro-production.
Fig.~\ref{fig:xsection} collects the forward
($x_{\rm F} > 0$)
$D$ meson production cross
sections as measured by several experiments, as a function of 
$\sqrt{s}$, the center of mass energy.  Where appropriate, the values
have been adjusted to 
reflect the current knowledge of branching ratios \cite{pdg96}.
The cross section values for proton induced reactions, published by
experiments NA27, E743 and E653 
for all \xf, were divided by 2.
The results obtained with nuclear targets were divided by the target
atomic mass number ~A.
The error bars reflect statistical and systematical
uncertainties added in quadrature and include the uncertainty on
the branching ratios.
The order of magnitude discrepancies found in earlier
measurements have disappeared and the energy dependence of the data
is rather consistent with PYTHIA based QCD predictions
(solid lines in Fig.~\ref{fig:xsection}).
The calculations do not reproduce very well the observed $D^+/D^0$ ratio 
and thus two different up-scale factors $K$ would be needed to reproduce
charged and neutral $D$ cross sections simultaneously.
(See also a comment concerning this situation in \cite{Mangano}).
Since the spectral distributions for $D^+$ and $D^0$
($D^-$ and $\overline{D}^0$)
are similar (see below), we use in the following an average $K$ factor
of 5.0 determined to reproduce the overall $D$ yield.

The hadro-production of open charm mesons on nuclear targets has
recently been precisely measured in collisions induced by 800~GeV
protons.  Parameterizing the nuclear target dependence in the usual
way, $\sigma(pA\to D+X) = \sigma_N A^\alpha$, with $\sigma_N$ the open
charm cross section for proton-nucleon collisions,
the E789 experiment has obtained
$\alpha=1.02\pm0.03\pm0.02$, for $0.0<$\xf$<0.08$ (average \xf\ of
0.031), comparing Be and Au targets~\cite{e789}.
With 250~GeV $\pi^+$ and $\pi^-$
beams, the E769 experiment has measured
$\alpha=1.00\pm0.05\pm0.02$ for $D$ mesons with \xf\ between 0.0 and
0.8, using a target system composed of Be, Al, Cu and W foils~\cite{Alves93}.
Besides, E769 has not found any significant dependence of $\alpha$ on
\pt\ or \xf, at least for the \xf$<0.5$ region, responsible for most
of the cross section.
These findings confirm what one expects for a hard process, in which
the partons in the nuclear target independently contribute to the total
cross section.

In the following we will use hadron-hadron and hadron-nucleus data,
along with the atomic mass number dependence, to evaluate to which extent
charm production is described by leading order QCD calculations.
To this end we have performed calculations with the code \Pythia\ using
MRS~G parton distributions \cite{pdf}
and \mkt=1~GeV$^2$/$c^2$.
We checked that
using \mkt=2~GeV$^2$/$c^2$
changes the dilepton yield by less than 10\%. Most of the calculations
were performed for proton-proton collisions. We have, however, checked
that isospin dependences are small for open charm production. At
$\sqrt{s}$ = 18 \gev\ the differences predicted by \Pythia\ in D-meson
distributions from 
proton-proton and neutron-neutron collisions are below 10 \%. Since
open charm production is mostly due to gluon fusion this is not unexpected.

\begin{figure}[htb]
\centering
\resizebox{0.8\textwidth}{!}{\includegraphics{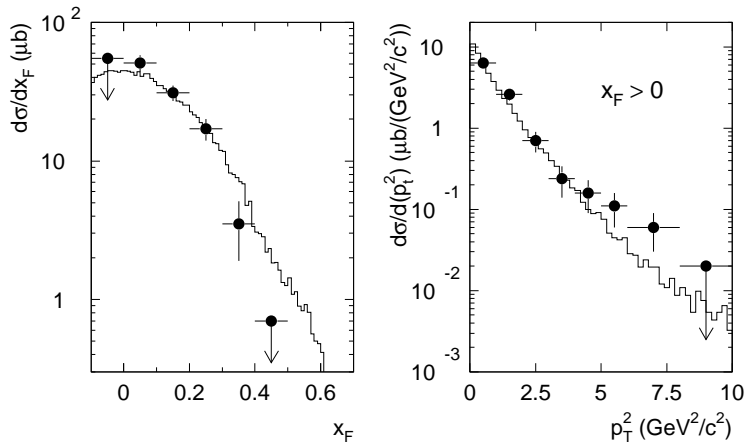}}
\caption{{}$D^+,D^-,D^0,\overline{D}^0,D^+_s,D^-_s$ production in p-p
  at 250~GeV: comparison between experiment \cite{alv96b} (black dots) and
  the up-scaled \Pythia\ calculation (histograms).}
\label{fig:xfpt}
\end{figure}
The energy dependence of charm production is rather well reproduced
by these calculations (Fig.~\ref{fig:xsection}).
In Fig.~2 we present a comparison of \Pythia\
predictions for \pt\ and \xf\ distributions with data from
the E769 collaboration at a beam energy of 250~GeV~\cite{alv96b}.
The absolute
normalization of the calculations has been adjusted to reproduce the
$D$ meson data ($D^+$, $D^-$, $D^0$, $\overline{D}^0$, $D^+_s$, $D^-_s$),
yielding an average $K$ factor of 5.0, which is used for
all subsequent calculations, independent of beam momentum.  Notice
that the value of $K$ depends on the $c$-quark mass used;
here we have used $m_c=1.35$\gevcc.
It also depends on the choice of the $Q^2$ scale used in \Pythia.  In
our calculations we have used $Q^2=$\emph{\^s}, the parton-parton
center of mass energy squared.  We have verified that, if the $Q^2$
scale is given by \emph{\^m}$_{T}^2$ (\Pythia 's default), the calculated cross sections 
increase by a factor of two, implying a corresponding decrease in the
$K$ factor. Here
\emph{\^m}$_{T}$ is the transverse mass of the outgoing particles
produced in the hard scattering.
Note that the kinematical distributions do not depend appreciably on the
choice of the $Q^2$ scale.

Apart from the absolute normalization, the \Pythia\
calculations reproduce reasonably well the measured \pt\ and \xf\ dependence,
implying that the basic kinematics of the charm production process is
correctly simulated. 

The agreement with the data points can be further improved
by using a $\delta$-function for the fragmentation of the 
charmed quarks in the charmed mesons.  However, we find it 
difficult to believe that a $\delta$-function can be a good 
\begin{figure}[hb]
\centering
\resizebox{0.8\textwidth}{!}{\includegraphics{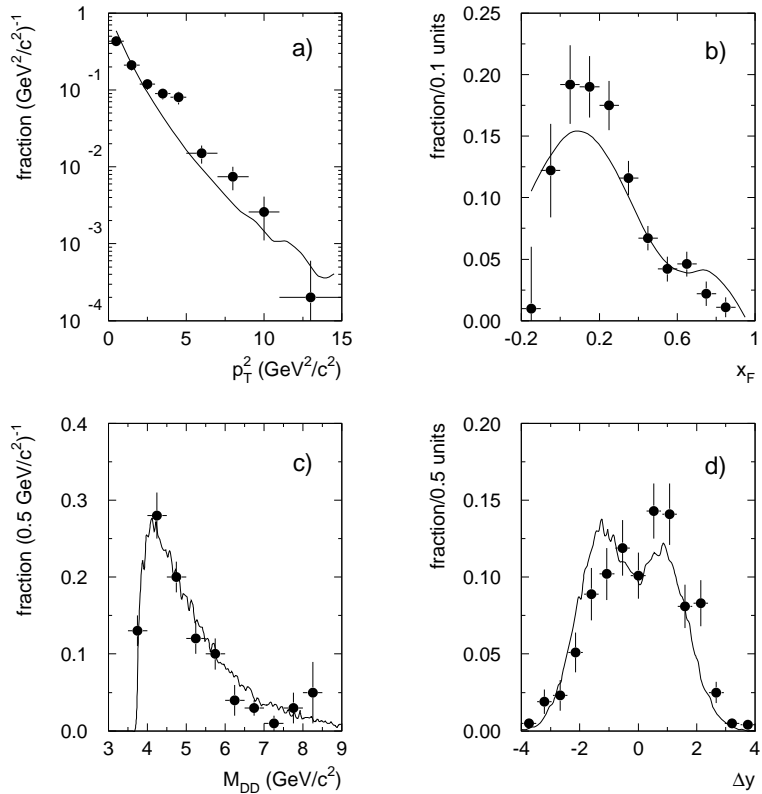}}
\caption{{}
   $D\overline{D}$ data from $\pi^-$-Cu collisions at
   {\protect $\protect \sqrt{s}\protect $}=26 GeV \cite{wa92}.
   Parts a) and b) represent the $p_{\rm T}^2$ and {\xf} distributions of the
   $D\overline{D}$ pair.
   Parts c) and d) show the $D\overline{D}$ invariant
   mass and the rapidity difference $\Delta y=y_D-y_{\overline{D}} \ $
   distributions.
   The data are compared to the \Pythia\ calculation
   (solid line).  The \Pythia\ distributions are normalized
   in the same way as the data, i.e.\ such that
   the integral over the displayed range is unity.}
\label{fig:kine}
\end{figure}
description of the hadronization process and, therefore,
have used, in the calculations reported in this paper, the
Lund string fragmentation model (\Pythia 's default).

To further check the kinematics of the charm production process as implemented
in \Pythia\, we present in Fig.~3 comparisons with the
WA92 data for \pim--~Cu interactions, where kinematical correlations
between charmed particles were measured at \rs=26~GeV~\cite{wa92}.
Fig.~3a shows good agreement between the transverse momentum
spectrum calculated with \Pythia\ and the data.
As shown in Figs.~3b, 3c, and 3d, also the Feynman x
distribution of the \ddbar\ pair, the distribution in \ddbar\
invariant mass $M_{D\overline{D}}$, as well as the rapidity difference
between $D^+$ ($D^0$) 
and $D^-$ ($\overline{D}^0$) are reasonably well reproduced by the \Pythia\
calculations, lending considerable confidence to the accuracy of the
calculated lepton pair mass spectra.

\begin{figure}[htb]
\centering
\resizebox{0.8\textwidth}{!}{\includegraphics{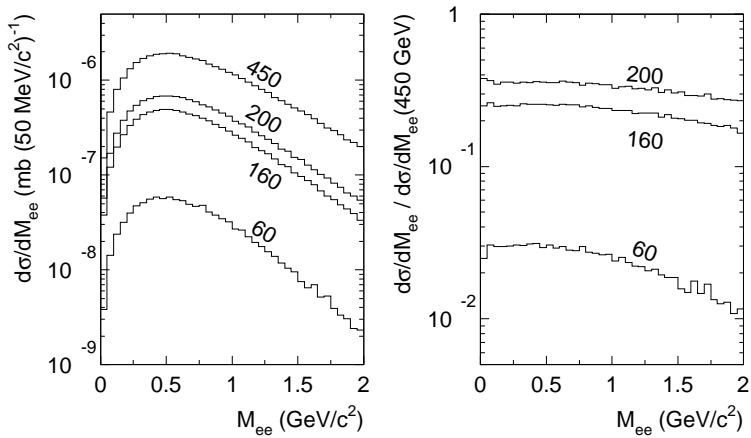}}
\caption{Left: Invariant mass spectra of dielectrons from charm decay
  in \pp\ collisions for different beam momenta (\Pythia).  Right: The
  same normalized to the spectrum at 450\gev.}
\end{figure}

To complete the systematics, and to lead into the discussion of charm
contributions to dilepton spectra, we present in Fig.~4 the
di-electron mass distributions from semi-leptonic $D$-decays, without
any phase space selection,
for various beam energies. The calculated mass distributions are all
rather broad, with a shallow maximum near $M_{\ee}=0.5$\gevcc. No
strong energy dependence is observed.
To emphasize this point we show,
in the right hand part of Fig.~4, the ratio of mass
distributions relative to the one computed at 450\gev. From this plot
one observes only a slight narrowing of the mass distributions with
decreasing beam momentum.
Especially in the mass range $0 < M_{\ee} <
1$\gevcc, the variation of the ratio with mass is less than 20\,\%.

\begin{figure}[htb]
\centering
\resizebox{0.7\textwidth}{!}{\includegraphics{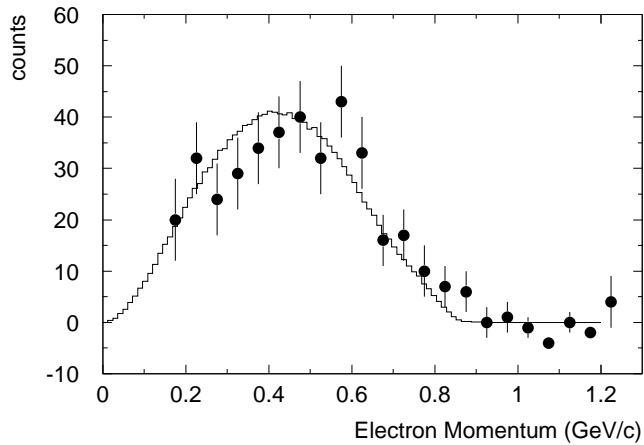}}
\caption{{}
Momentum distribution of electrons from $D^+$ and $D^0$ decays, 
measured in the rest frame of the decaying mesons. 
The experimental data (full dots) were taken from ref.~\cite{bac79}. 
The solid histogram was extracted from the \Pythia\ calculation. }
\label{D-decay}
\end{figure}

The shape of the lepton pair mass distributions depends 
crucially on the implementation in \Pythia\ of semi-leptonic $D$ decays.
In order to check if such  decays are treated correctly we compare,
in Fig.~\ref{D-decay},  the momentum distribution of electrons 
from $D$ decays in \Pythia\ with experimental distributions \cite{bac79}. 
The theoretical distributions, where modes with K mesons and with K$^*$
mesons contribute with roughly equal weight, are in good agreement with
the measured spectrum.  Furthermore, we have verified that the values
used in \Pythia\ for 
the semileptonic decay branching ratios are in excellent agreement 
with the most recent values of \cite{pdg96}.
We conclude, therefore, that uncertainties in
the lepton energy spectra and branching ratios from $D$ meson decays can
be neglected in the following discussion.

\begin{figure}[htb]
\centering
\resizebox{0.75\textwidth}{!}{\includegraphics{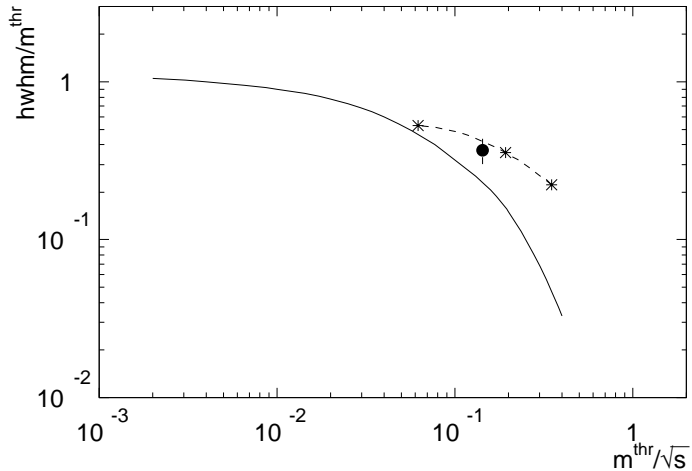}}
\caption{Relation between the collision energy and the half width at half
  maximum of the
  $M_{D\overline{D}}$ distribution, as proposed by Fischer
  and Geist (solid line) and as calculated by
  \Pythia\ for \pp\ (stars connected by dashed line).
  Black dot: WA92 measurement~\protect{\cite{wa92}}.
}
\end{figure}

The shape of the dielectron mass spectrum also depends on the
shape of the \ddbar\ invariant mass spectrum. In 1983, Fischer and
Geist~\cite{fis83} proposed a scaling curve for the ratio between the
width of the parent mass distribution and the threshold mass $m^{\rm
thr}=2 m_{\rm D}$, plotted against the ratio between threshold mass
and \rs. Here, $m_{\rm D}$ is the mass of the $D$ meson. This curve was
motivated by ISR data and by early QCD 
calculations\cite{com79}.  In Fig.~6 we compare it with the present
\Pythia\ calculations.  Although a trend similar to the Fischer--Geist
curve emerges from the present calculations, the detailed shape looks
quite different.  We note, however, that the WA92 measurement (black
dot in Fig.~6) is in much better agreement with the present calculations
than with the Fischer--Geist scaling.

\section{Dilepton yield from associated charm production in proton induced
  collisions}

In the previous section we have shown that calculations with the code
\Pythia\ describe reasonably well
the most recent data on $x_F$ and \pt\ distributions of
$D$ mesons as well as rapidity correlations and
invariant mass spectra of the \ddbar\ pairs
\cite{alv96b,wa92} in the center-of-mass energy regime of
$\sqrt{s}\sim20$~GeV. In addition, the characteristics of the
semi-leptonic decays of $D$ mesons  are correctly reproduced. This
gives us confidence that the level 
of lepton pair yield from charm production can be reliably
calculated for \pp\ and \pa\ collisions on the basis of a \Pythia\
simulation. We will now compare our estimate of the lepton pair yield
from charm production to data on the inclusive lepton pair
yield. Both leptons from $D$ meson and
from $\Lambda_c$ decays are included in the calculations.

\begin{figure}[hb]
\centering
\resizebox{0.6\textwidth}{!}{\includegraphics{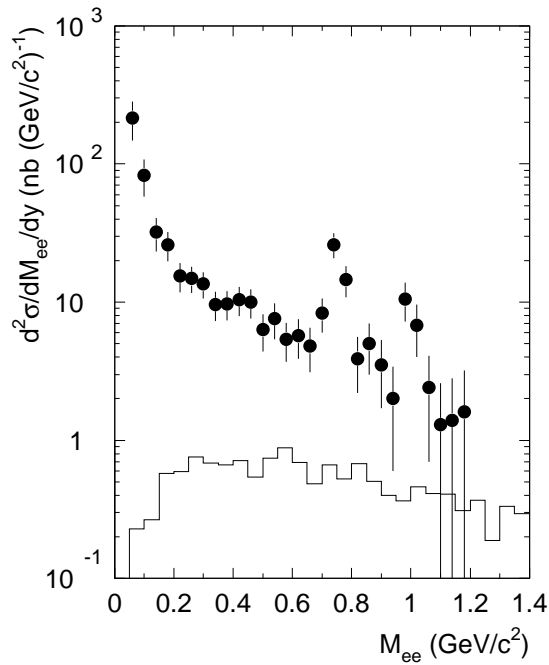}}
\caption{{}
  Invariant mass distribution for \ee\ pairs from \pp\ collisions
  at {\protect $ \protect \sqrt{s} \protect$}=60 GeV, measured at mid-rapidity. The solid line represents the
  \ee\ pairs from charm decays, as predicted by our \Pythia\ calculation.}
\label{fig:ISR}
\end{figure}

First we discuss the results of Chilingarov {\it et al}. \cite{chi79} who have
measured \ee\ pairs in \pp\ collisions at the ISR ($\sqrt{s} \approx 60$
GeV). For this energy \Pythia\ (including the above discussed $K$ factor)
predicts a 
\cc\ cross section of 100 $\mu$b, roughly in agreement with
the value
discussed in \cite{chi79}.  The \ee\ data of Chilingarov {\it et al.} and
our calculation for \ee\ pairs from open charm  are shown in
Fig.~\ref{fig:ISR}.
We have applied the cut 
\pt~$>1.6+0.5~M_{\ee}$ (with \pt\ and $M_{\ee}$ in GeV)
on the electron pair transverse momentum in order to simulate the experimental cuts
(see Fig.~5 in~\cite{chi79}).

According to the present calculations, and contrary to the conclusions
of  \cite{chi79},
the contribution
from charm decays is approximately one order of magnitude below the
observed dilepton yield while the shape of our computed \ee\ mass
spectrum is  rather similar
to that displayed in Fig.~6 of ref.~\cite{chi79}.
In order to check whether this result depends on the value of
\mkt\  used in the simulation,
we repeated the procedure with \mkt=2~GeV$^2$/$c^2$.
The corresponding yield was higher by only about 30\%. Of similar size
are the uncertainties due to the (small) differences between calculated
and measured \xf\ distributions for $D$ mesons (see Fig.~2). We assume that the
difference between our results and the conclusions of Chilingarov {\it
et al} is connected with the fact that, in ref. \cite{chi79}, the yield
for \ee\ 
pairs was calculated assuming isotropic decay in the pair rest frame,
while \Pythia\ incorporates  a 
strong anisotropy. In addition, in ref. \cite{chi79}, it was assumed
that the lepton pairs are produced with uniform rapidity
distribution. The fact that \Pythia\ reproduces well the observed
distribution of
rapidity gaps between the \ddbar\ mesons (see Fig.~3), which is at the root
of the anisotropy, lends strong support to the validity of the present
calculations. 
We therefore conclude that, at ISR energies,  the charm decay contribution to
the dilepton continuum below the $\phi$-mass is small. To explain the
observed continuum one should probably revisit the Dalitz decay
contributions to the mass spectra.

\begin{figure}[hb]
\centering
\resizebox{0.75\textwidth}{!}{\includegraphics{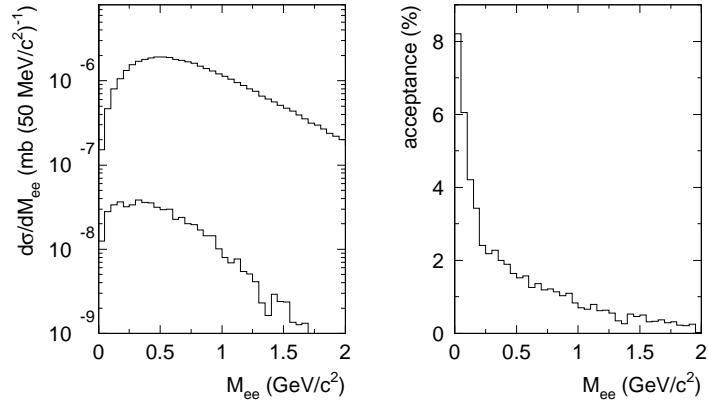}}
\caption{{} Left: Invariant mass spectra of dielectrons from charm decays
  in p-p collisions at beam energy 450\gev\ (\Pythia), before (upper
  histogram) and after (lower histogram) applying the CERES filter.
  Right: the ratio of the two histograms.
}
\label{fig:cuts}
\end{figure}
Recently the CERES/NA45 experiment has measured \ee\ pairs in \mbox{p-Be}
and \mbox{p-Au} collisions at 450 \gevc\ incident proton momentum \cite{aga95}.
In order to
compare our \Pythia\ calculation to the CERES data, we included in our
simulation their momentum resolution and kinematic cuts (\pt~$>50$~MeV/$c$
and $2.1<\eta<2.65$ for electrons and positrons).
The effect of the CERES acceptance on the \ee\ mass spectra
from charm decays is severe, as can be seen from Fig.~\ref{fig:cuts},
where we have compared  \ee\ mass distributions from charm before and after
applying the acceptance cut.
The integral of the mass distribution is
reduced by approximately a factor of 60 and, in addition, the shape is
distorted, leading to a steeper mass dependence for large pair masses.
The  dependence of the cross section on the nuclear mass number was taken
into account as outlined in the next section. Finally the
cross section was converted into a multiplicity distribution  and
scaled by $
<{\rm d}N_{ch}/{\rm d}y>^{-1}$ taken from the CERES experiment 
which gives the same normalization as used by the CERES collaboration.

\begin{figure}[htbp]
\centering
\resizebox{0.9\textwidth}{!}{\includegraphics{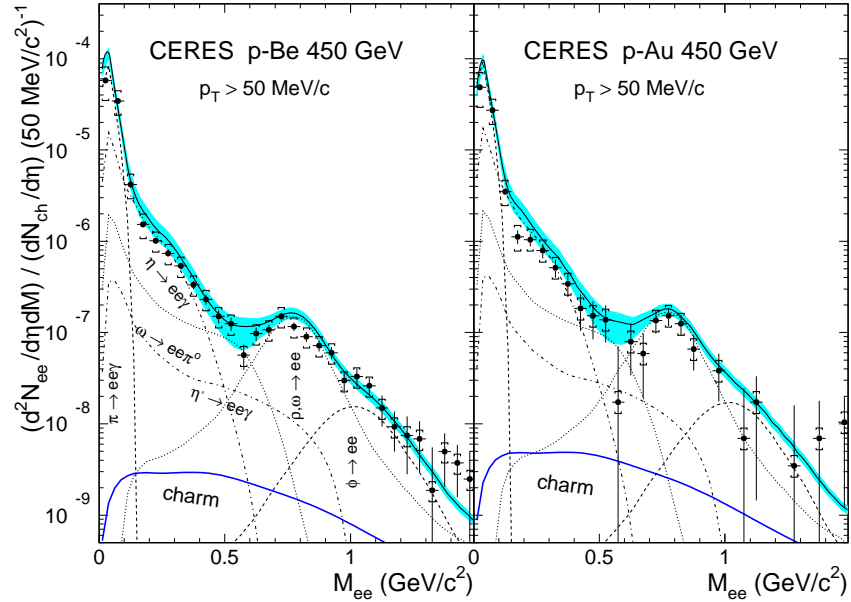}}
\caption{{} Invariant $e^+e^-$  mass distribution from 450 \gevc\ p-Be
and p-Au collisions measured by the CERES experiment. The background
from associated charm production has been added
to the hadron decay contribution according to our estimate.}
\label{fig:ceresp}
\end{figure}

Fig.~\ref{fig:ceresp} shows the CERES data as published in
ref. \cite{aga95}, but including our estimate of the contribution from
charm production. Clearly, also in p-nucleus collisions charm production
does not give an important contribution to the dilepton continuum below
the light vector mesons.

HELIOS/NA34-1 \cite{ake95} has published a measurement of \ee\
and \mm\  pairs in p-Be collisions at 450 \gevc\ momentum for
the same mass region.
From their analysis one concludes that the charm decay contribution
to their data is negligible.

While lepton pairs from associated charm production can be neglected
for low invariant masses, for masses above the
$\phi$-meson the situation is quite different.
Two experiments -- NA38 and HELIOS/NA34-3 -- have measured muon pairs from
p-nucleus collisions in this mass range. Both experiments have shown that
the dimuon cross section in p-W collisions at 200 \gevc\ can be
accounted for by charm production and pairs produced via the Drell-Yan
mechanism. The HELIOS-3 estimate \cite{mas95} of the charm cross section was
based on an approach similar to the one presented in this paper.
The experiment NA38
used \Pythia\ to determine the shape of the dilepton background from charm
production but fitted the charm cross section to their muon pair
data \cite{clqm93}.
The result is compatible with the cross sections shown in
Fig.~\ref{fig:xsection}.
Both experiments consistently show that, for masses above the
$\phi$ meson, $c\bar{c}$-production is an important source of
lepton pairs until the continuum is overwhelmed by pair production via the
Drell-Yan mechanism in the mass region beyond the \jpsi.

\section{Extrapolating to nucleus-nucleus collisions}

\begin{figure}[htb]
\centering
\resizebox{0.45\textwidth}{!}{\includegraphics{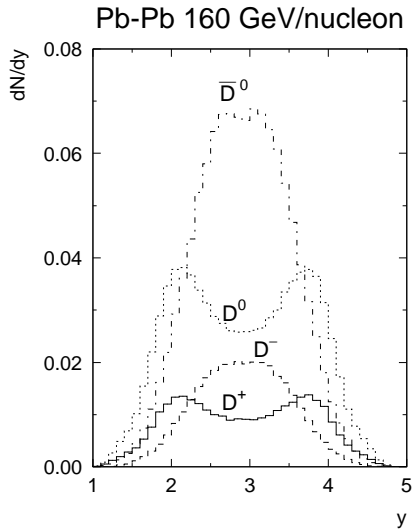}}
\caption{{} Calculated rapidity distributions of
  $D^+,D^-,D^0,\overline{D}^0$ from central (5\,\%) Pb+Pb collisions at
  160~\gevc/nucleon.  The ordinate represents absolute multiplicities
  per unit rapidity.}
\label{fig:dmesons}
\end{figure}

Since the charm cross section scales with the number of target nucleons
going from p-p to p-nucleus collisions, we scale the p-p calculations
from \Pythia\ with the total number of nucleon-nucleon collisions to
obtain the nucleus-nucleus cross section. As discussed above, isospin
effects can be neglected for the 
calculation of open charm production within an
accuracy 
of better than 10 \%. The
total number of 
nucleon-nucleon collisions is calculated from the nuclear collision
geometry based on the concept of an overlap
function $T_{AB} (b)$ for $A$ + $B$ collisions at impact
parameter $b$ \cite{esk89}.
This leads to:
\beq \frac{\dd\sigma}{\dd y}^{AB} =
\frac{\dd\sigma}{\dd y}^{pp} \int_{b_1}^{b_2}\dd^2b ~ T_{AB}(b).
\label{sigab}
\eeq
Here, the integration limits indicate the impact parameter range
covered in the experiment. For minimum bias collisions, where
$b_1=0$ and $b_2 = \infty$ this leads to
\beq
\frac{\dd\sigma}{\dd y}^{AB} =  A B ~ \frac{\dd\sigma}{\dd y}^{pp}
\label{sigabmb}
\eeq
For the rapidity density distributions this implies
\beq
\frac{\dd N}{\dd y}^{AB} = \frac{\dd N}{\dd y}^{pp}~\sigma_{inel}^{pp}~
\frac{\int_{b_1}^{b_2}d^2b
~T_{AB}(b)}{\pi(b_2^2-b_1^2)}.
\label{rap}
\eeq
Here, $\sigma_{inel}^{pp}$ is the inelastic nucleon-nucleon cross
section at the appropriate energy\,\footnote{\,In actual calculations the
inelastic nucleon-nucleon cross section drops out as we compute directly
$\frac{\dd\sigma}{\dd y}^{pp}= \frac{\dd N}{\dd
y}^{pp}~\sigma_{inel}^{pp}$.}. Equation \ref{rap}, when applied 
to central collisions ($b\approx 0$) and using a sharp sphere
approximation of the nuclear density profile
for the evaluation of the overlap function, results in
the familiar dependence
\beq \frac{\dd N}{\dd y}^{AA}(b\sim 0) \approx
\frac{\dd N}{\dd y}^{pp}~ A^{4/3}.
\label{rapAA}
\eeq
We note that this approach neglects possible effects due to the Fermi
motion of the nucleons in the colliding nuclei. However, the threshold
for open charm production is \rs~=~5.5~GeV compared to the \rs~=~17.3~GeV
available in Pb-Au collisions so that Fermi motion effects are not
expected to play an important role.
We use Woods-Saxon distributions for
the nuclear densities \cite{saxon} to evaluate the thickness function
T$_{\rm AB}{\rm (b)}$ (or T$_{\rm A}$ for the case of \pa\ collisions) and adjust
the impact parameter range to obtain the fraction of
inelastic cross section sampled in each experiment.
Applying this procedure to the \pp\ results from \Pythia\
we obtain the $D$ meson yield expected for nucleus-nucleus collisions.
In Fig.~\ref{fig:dmesons} we show the prediction for the rapidity
distribution of $D$ mesons in central (5\,\% of geometrical cross section)
Pb+Pb collisions at 160~\gevc/nucleon. The distributions exhibit rather
drastic differences between $D$ mesons and their anti-particles.

These differences originate, within the \Pythia\ framework, from the
interaction of $c$ and $\overline{c}$  
quarks with valence quarks of the colliding nucleons which takes place prior to
hadronization.  
While the $\overline{c}$ forms a string with a single valence quark, 
the $c$ interacts with valence diquarks, and is  dragged towards 
beam or target rapidity. 
As a consequence, the $D^+$ and $D^0$ meson rapidity distributions are 
broadened and, at  energies around $\sqrt{s}=$20 \gev, even exhibit minima at
midrapidity.  

Similar asymmetries, although different in detail because of the
different quark content of the projectile, have recently been observed
for $\pi^-$ - nucleus
interactions at 500~\gevc\ \cite{e791}. There it is also shown how well
these asymmetries are reproduced by \Pythia\,\footnote{\,In
ref. \cite{e791} it is shown that the measured asymmetries can be
well reproduced by \Pythia\ if one sets the mass of the charmed quark to
1.7 \gevcc\ and otherwise uses the same parameters as are used in the
present investigation. With our parameter set the agreement between data
and calculations is only slightly deteriorated.}.
It would be interesting to verify experimentally whether such
asymmetries also exist in proton induced collisions, in particular at
\rs\ $\sim20$~\gev, since according to \Pythia\ these asymmetries
tend to disappear at higher beam energies.
 
The shapes of the $D$ meson rapidity distributions for \PbPb\
collisions might also be distorted in the nuclear environment, as
observed for \pt\ distributions measured in p-nucleus collisions, commonly
referred to as Cronin effect~\cite{cronin}. Clearly, a direct
measurement of $D$ 
meson production in \PbPb\ collisions would be of great interest.  The
possible modifications discussed above should not change the total
rate significantly and we expect that (in the absence of charm
enhancement) the rapidity density of $D$ mesons from \PbPb\ collisions
at 160~GeV/nucleon is ${\rm d}N/{\rm d}y$=0.10--0.15 at midrapidity.

\begin{figure}[htp]
\centering
\resizebox{0.55\textwidth}{!}{\includegraphics{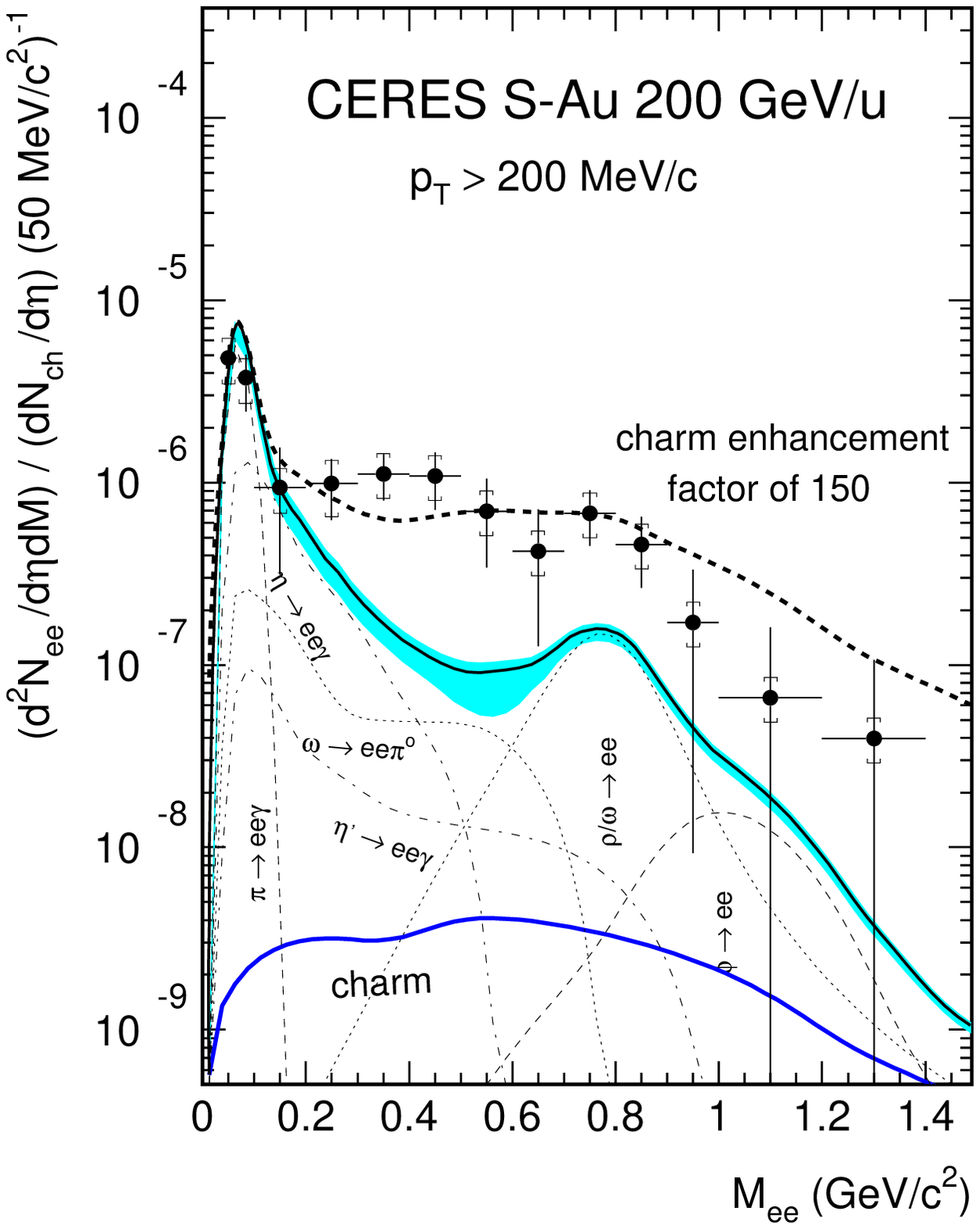}}
\caption{{} Invariant mass spectrum of \ee\ pairs measured by CERES
  in 200 GeV/nucleon S-Au collisions. The solid line gives the
  contribution expected from charm production in the standard scenario.
The dashed line results from adding to the hadron decay background a
  charm contribution enhanced by a factor of 150. }
\label{fig:ceress}
\bigskip
\centering
\resizebox{0.8\textwidth}{!}{\includegraphics{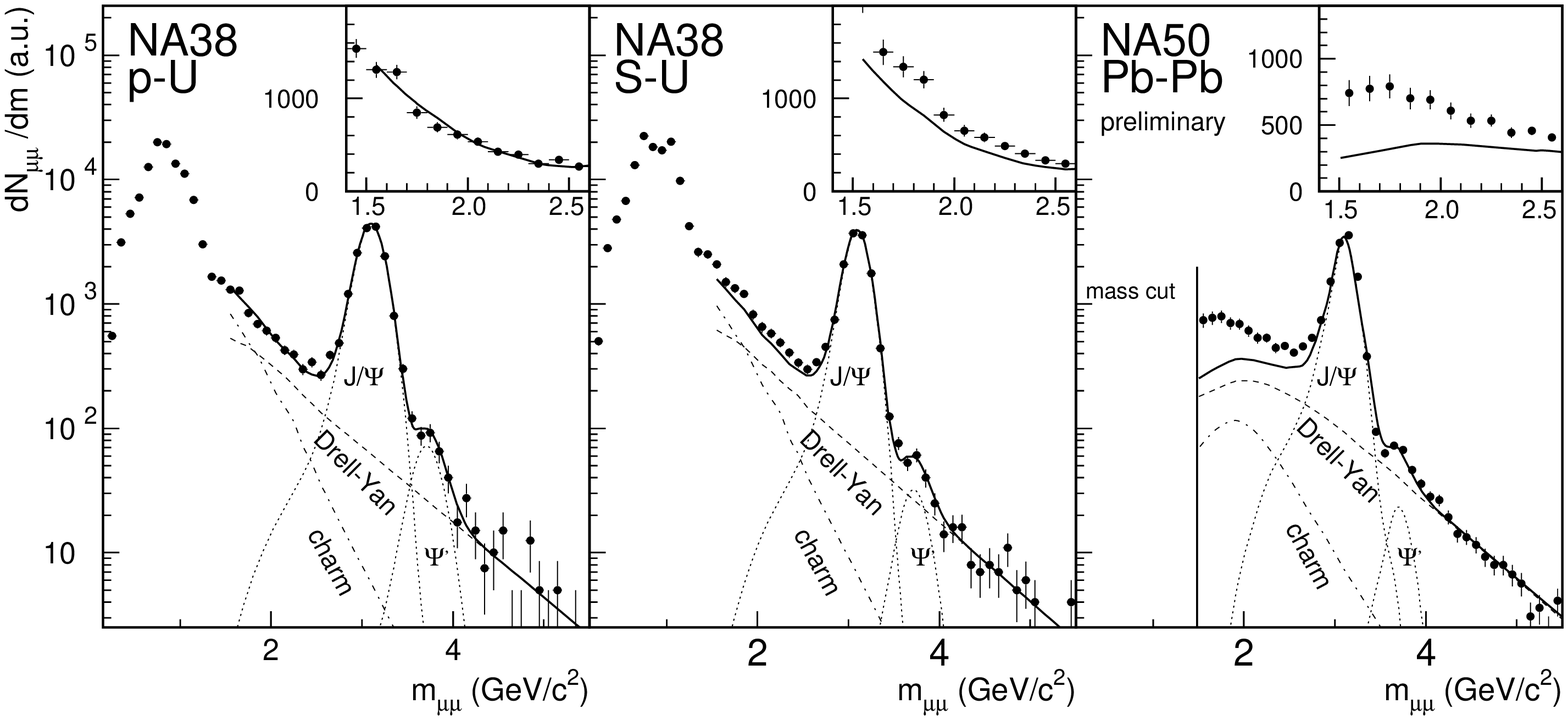}}
\caption{Dimuon spectra from p-U, S-U, and Pb-Pb collisions
(NA38/NA50 Collaboration).
The S-U spectrum excludes charm enhancement by more than a factor of 3.
The figure was taken from \cite{dre96} and is based on results from
\cite{sco96}.
}
\label{fig:na38}
\end{figure}

As in the previous section we want to focus on the charm decay contribution
to the inclusive lepton pair production. In Fig.~\ref{fig:ceress} the
results of our calculation are compared to the S-Au \ee\ data
published by the CERES group \cite{aga95}.
Clearly, charm production in the ``standard''
scenario  as presented here contributes less than one percent of the observed
dilepton yield. The slight difference in shape of the charm
contribution compared to Fig.~\ref{fig:ceresp} is solely due to the higher
\pt-cut used by the CERES experiment for S-Au collisions.

It is not unreasonable \cite{wong} to contemplate the possibility of
a strong additional source of charm in nucleus-nucleus collisions.
However, an enhancement by a factor of 150 would be  needed to describe the
present CERES data.
The result of such an  enhancement is shown as the dashed line
in Fig.~\ref{fig:ceress}.
Though it roughly explains the overall
dilepton yield, it predicts a flatter mass distribution than is
observed  resulting in an overestimate of the yield at high
masses.
The results for dimuon production obtained by  the NA38/NA50 Collaboration are
shown in Fig.~\ref{fig:na38}
where also the predictions for the charm and Drell-Yan continuum are
indicated.
Based on these results it is clear that charm enhancement factors larger
than 3 would lead to inconsistencies with the S-U dimuon data.

\section{Conclusions}

In summary, the level of charm production in p-p and p-A collisions in
the energy range \rs~$\sim 20-40$~GeV is now reasonably well
measured. The trends in the data are well reproduced by leading
order perturbative QCD calculations performed using the code  \Pythia\, with
\mkt=1~GeV$^2$/$c^2$. The
calculations were extrapolated to nucleus-nucleus collisions by scaling
with the geometrically allowed number of nucleon-nucleon collisions. We
have used these calculations to determine the dilepton pair contribution
expected 
from associated $c\bar{c}$ production. In the mass region below the
vector mesons $\rho, \omega$ and $\phi$ this contribution is negligible
compared to the yield arising from decays of light mesons for all
collision systems investigated in this paper.
For S-Au collisions the prediction is two orders of magnitude below
the experimentally observed \ee\ continuum.
Since any possible enhancement of charm production in
nuclear collisions is limited by \mm\ data in the mass region
above the $\phi$ to less than a factor of 3, any explanation of the
observed anomalous low mass pair production based on charm production
can be ruled out.

We thank George Ginther, Karl Harrison and Thomas Ullrich
for useful discussions while collecting the charm cross section
measurements from different experiments.
We are grateful to H.~G.~Fischer for helpful and provocative remarks
about charm production and to W.~Geist, M.~Mangano and A.~Morsch for
helpful discussions.

\end{document}